\documentclass[a4paper]{amsart}

\usepackage{amssymb,enumerate,amsmath}

\usepackage[utf8x]{inputenc}
\usepackage[english]{babel}
\usepackage[T1]{fontenc}
\usepackage[left=2cm,right=2cm,top=2cm,bottom=2cm]{geometry}

\usepackage{tikz}
\usepackage{mathtools}
\usepackage[colorlinks]{hyperref}
\usepackage{mathrsfs} 
\usepackage{faktor}
\usepackage{fancyvrb}
\RecustomVerbatimEnvironment{Verbatim}{Verbatim}{formatcom=\footnotesize,baselinestretch=0.6,xleftmargin=5mm,xrightmargin=5mm}

\newcommand{\up}[1]{\mathrm{e}^{\text{\tiny$+$}}_{#1}}
\newcommand{\down}[1]{\mathrm{e}^{\text{\tiny$-$}}_{#1}}
\newcommand{\ZZ}{\mathbb{Z}}
\newcommand{\KK}{\mathbb{K}} 
\newcommand{\xx}{\boldsymbol{x}}
\newcommand{\al}{\boldsymbol{\alpha}}

\renewcommand{\geq}{\geqslant}
\renewcommand{\leq}{\leqslant}

\newcommand{\QQ}{\mathbb{Q}}

\newcommand{\TT}{\mathbb{T}}

\newtheorem{theorem}{Theorem}[section]

\newtheorem{proposition}[theorem]{Proposition}

\newtheorem*{gotzmannDec}{Gotzmann's decomposition}
\newtheorem*{gotzmannReg}{Gotzmann's Regularity Theorem}

\theoremstyle{definition}
\newtheorem{definition}[theorem]{Definition}

\theoremstyle{remark}
\newtheorem{example}[theorem]{Example}

\begin{document}

\title{Strongly Stable Ideals and Hilbert polynomials}

\author[D.~Alberelli]{Davide Alberelli}

\address{Davide Alberelli}
\email{\href{mailto:davide.alberelli@gmail.com}{davide.alberelli@gmail.com}}

\author[P.~Lella]{Paolo Lella}
\address{Paolo Lella\\ Dipartimento di Matematica\\ Politecnico di Milano\\ 
         Piazza Leonardo da Vinci 32\\ 20133 Milano\\ Italy.}
\email{\href{mailto:paolo.lella@polimi.it}{paolo.lella@polimi.it}}
\urladdr{\url{http://www.paololella.it/}}

\thanks{The second author is a member of GNSAGA}
\subjclass[2010]{13P10, 13P99}

\begin{abstract}
The \texttt{StronglyStableIdeals} package for \textit{Macaulay2} provides a method to compute all saturated strongly stable ideals in a given polynomial ring with a fixed Hilbert polynomial. A description of the main method and auxiliary tools is given. 
\end{abstract}

\keywords{Strongly stable ideal, Borel-fixed ideal, Hilbert polynomial, Gotzmann number, Hilbert scheme}

\maketitle

\section*{Introduction} 

Strongly stable ideals are a key tool in commutative algebra and algebraic geometry. These ideals have nice combinatorial properties that make them well suited for both theoretical and computational applications. In the case of polynomial rings with coefficients in a field of characteristic zero, the notion of strongly stable ideals coincides with the notion of Borel-fixed ideals. Such ideals are fixed by the action of the Borel subgroup of triangular matrices and play a special role in theory of Gr\"obner bases because initial ideals in generic coordinates are of this type \cite{Galligo}.

In the context of parameter spaces of algebraic varieties, Galligo's theorem says that each component and each intersection of components of a Hilbert scheme contains at least a point corresponding to a scheme defined by a Borel-fixed ideal. Hence, these ideals are distributed throughout the Hilbert scheme and can be used to study its local structure. To this aim, in recent years several authors \cite{LellaRoggero,CioffiRoggero,BCLR,LR2,BCR-GG,BCR} developed algorithmic methods based on the use of strongly stable ideals to construct flat families corresponding to special loci of the Hilbert scheme. In particular, a new open cover of the Hilbert scheme has been defined using strongly stable ideals and the action of the projective linear group \cite{BLR,BLMR}. In this construction, the list of all points corresponding to Borel-fixed ideals in a given Hilbert scheme is needed. The main feature of the package \texttt{StronglyStableIdeals} is a method to compute this set of points, i.e.~the list of all saturated strongly stable ideals in a polynomial ring with a given Hilbert polynomial. The method has been theoretically introduced in \cite{CLMR} and improved in \cite{LellaBorel}. Several other tools are developed and presented in the paper.

\section{Strongly stable ideals} 

Let us denote by $\KK[\boldsymbol{x}]$ the polynomial ring in $n+1$ variables $\KK[x_0,\ldots,x_n]$ with coefficients in a field $\KK$. We assume $x_0 > x_1 > \cdots > x_n$. We use the multi-index notation to describe monomials, i.e.~for every $\boldsymbol{\alpha} = (\alpha_0,\ldots,\alpha_n) \in \ZZ_{\geq 0}^{n+1}$, $\xx^{\al} := x_0^{\alpha_0}\cdots x_n^{\alpha_n}$ and we denote by $\TT_{n,s}$ the set of monomials of $\KK[\xx]$ of degree $s$. For any monomial $\xx^{\al}$, we denote by $\min \xx^{\al}$ and $\max \xx^{\al}$ the indices of minimal and maximal variable dividing $\xx^{\al}$. 

 Following \cite{GreenGIN}, \emph{increasing} and \emph{decreasing elementary moves} are defined as the multiplications
\begin{equation*}
\up{i}(\xx^{\al}) := \frac{x_{i-1}}{x_i} \cdot \xx^{\al},\ i > 0\qquad\text{and}\qquad \down{j}(\xx^{\al}) := \frac{x_{j+1}}{x_j}\cdot \xx^{\al},\ j < n.
\end{equation*}
We say that an elementary move $\mathrm{e}_i^{\text{\tiny +/$-$}}$ is \emph{admissible} for a monomial $\xx^{\al}$ if $\alpha_i > 0$, i.e.~$\mathrm{e}_i^{\text{\tiny +/$-$}}(\xx^{\al})$ is a monomial of $\KK[\xx]$.

\begin{definition} An ideal $I \subset \KK[\xx]$ is called \emph{strongly stable} if
\begin{enumerate}[(i)]
\item $I$ is a monomial ideal;
\item for every $\xx^{\al} \in I$ and for every admissible increasing move $\up{i}$, the monomial $\up{i}(\xx^{\al})$ is contained in $I$.
\end{enumerate}
\end{definition}

We recall that a strongly stable ideal is a Borel-fixed ideal. We now summarize some properties holding in general for Borel-fixed ideals and useful in this context.

\begin{proposition}[{\cite[Section 2]{GreenGIN}}]\label{prop:BorelProperties} Let $I \subset \KK[\xx]$ be a strongly stable ideal. 
\begin{enumerate}[(i)]
\item\label{it:BorelProperties_i} The regularity of $I$ is equal to the maximal degree of a generator.
\item\label{it:BorelProperties_ii} Let $\mathfrak{m}$ be the irrelevant ideal of $\KK[\xx]$. Then, $(I:\mathfrak{m}) = (I : x_n)$, so that the ideal $I$ is saturated if no generator involves the last variable $x_n$.
\item\label{it:BorelProperties_iii} The last variable $x_n$ is a regular element for $I$, i.e.~the multiplication by $x_n$ induces the short exact sequence
\[
0\ \longrightarrow\ \faktor{\KK[\xx]}{I}(t-1) \ \xrightarrow{\cdot x_n}\ \faktor{\KK[\xx]}{I}(t)\ \longrightarrow \faktor{\KK[\xx]}{(x_n,I)}(t)\ \longrightarrow\ 0.
\]
\end{enumerate}
\end{proposition}

\section{Hilbert polynomials}
The Hilbert polynomial $p(t)$ of a homogeneous ideal $I \subset \KK[\xx]$ is the numerical polynomial such that for $s$ sufficiently large
\[
\dim_\KK \left(\faktor{\KK[\xx]}{I}\right)_s = \dim_\KK \left(\faktor{\KK[\xx]_s}{I_s}\right) = \binom{n+s}{n} - \dim_{\KK} I_s = p(s).
\] 
Obviously, not every numerical polynomial is a Hilbert polynomial of some homogeneous ideal. Those being Hilbert polynomials have been completely described by Gotzmann \cite{Gotzmann}.

\begin{gotzmannDec}\label{th:gotzmannDec}
A numerical polynomial $p(t) \in \QQ[t]$ is a Hilbert polynomial if, and only if, it can be written as
\begin{equation}\label{eq:gotzmannDec}
p(t) = \binom{n+a_1}{a_1} + \binom{n+a_2-1}{a_2} + \ldots + \binom{n+a_r-(r-1)}{a_r},\qquad a_1 \geq a_2 \geq \cdots\geq a_r \geq 0.
\end{equation}
\end{gotzmannDec}

This decomposition is strictly related to Macaulay's decomposition
\[
p(t) = \sum_{k=0}^d \left[\binom{t+k}{k+1} - \binom{t+k-m_k}{k+1}\right]
\]
where $d = \deg p(t)$. For all $n\geq d+1$ the saturated lexicographic ideal in $\KK[x_0,\ldots,x_n]$ with Hilbert polynomial $p(t)$ is
\[
(x_0^{},\ldots,x_{n-d-2}^{},x_{n-d-1}^{b_d+1},x_{n-d-1}^{b_d} x_{n-d}^{b_{d-1}+1},\ldots,x_{n-d-1}^{b_d} x_{n-d}^{b_{d-1}} \cdots x_{n-1}^{b_0}),
\]
where
\[
b_d = \# \{ a_j\ \vert\ a_j = d \} = m_d \quad\text{and}\quad b_k = \# \{ a_j\ \vert\ a_j = k \} = m_k - m_{k+1},\ 0\leq k < d.
\]

The description of the lexicographic ideal in terms of Gotzmann's decomposition gives an insight of the following theorem.

\begin{gotzmannReg}
The regularity of a saturated ideal $I \subset \KK[\xx]$ with Hilbert polynomial $p(t)$ is at most $r$, where $r$ is the number of terms in the decomposition \eqref{eq:gotzmannDec} and it is called \emph{Gotzmann number} of $p(t)$.
\end{gotzmannReg}

\begin{example}
The package \emph{StronglyStableIdeals} provides the method \texttt{isHilbertPolynomial} to determine if a numerical polynomial is a Hilbert polynomial.

\smallskip

\begin{Verbatim}
Macaulay2, version 1.11
with packages: ConwayPolynomials, Elimination, IntegralClosure, InverseSystems, LLLBases,
               PrimaryDecomposition, ReesAlgebra, TangentCone

i1 : loadPackage "StronglyStableIdeals";

i2 : QQ[t];

i3 : isHilbertPolynomial (4*t)

o3 = true

i4 : isHilbertPolynomial (5*t-6)

o4 = false
\end{Verbatim}
Gotzmann's and Macaulay's decompositions of a Hilbert polynomial can be computed with the methods \texttt{gotzmannDecomposition} and \texttt{macaulayDecomposition}. These methods return the list of terms in the decompositions. The summand $\binom{t+e}{c}$ is constructed with the command \texttt{projectiveHilbertPolynomial(c,c-e)}.

\smallskip

\begin{Verbatim}
i5 : gotzmannDecomposition (4*t)

o5 = {P , - P  + P , - 2*P  + P , - 3*P  + P , P , P }
       1     0    1       0    1       0    1   0   0

o5 : List

i6 : macaulayDecomposition (4*t)

o6 = {- P  + P , 7*P  - P , - P  + P , - 10*P  + 5*P  - P }
         0    1     0    1     1    2        0      1    2

o6 : List
\end{Verbatim}
Finally, the saturated lexicographic ideal $L$ with Hilbert polynomial $p(t)$ in the polynomial ring $\KK[\xx]$ can be computed with the method \texttt{lexIdeal} and its regularity is equal to the Gotzmann number of $p(t)$.

\smallskip

\begin{Verbatim}
i7 : L = lexIdeal (4*t, QQ[x,y,z,w])

                5   4 2
o7 = ideal (x, y , y z )

o7 : Ideal of QQ[x, y, z, w]

i8 : regularity L == gotzmannNumber (4*t)

o8 = true
\end{Verbatim}
\end{example}

\section{The main algorithm}
In this section, we outline the strategy of the main algorithm. This algorithm was firstly described in \cite{CLMR} and then optimized in \cite{LellaBorel}. The same problem has been previously discussed in \cite{Reeves} and an alternative algorithm was later presented in \cite{MN}.

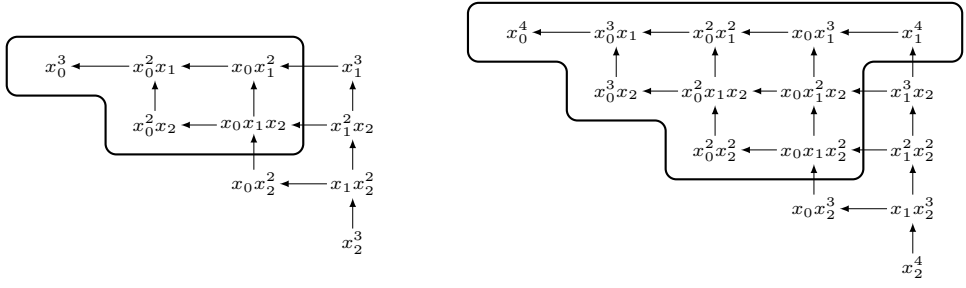
\begin{figure}
\begin{center}
\begin{tikzpicture}[>=latex,scale=0.65]

\node (300) at (0,0) [inner sep = 1pt] {\scriptsize $x_0^3$};
\node (210) at (2,0) [inner sep = 1pt] {\scriptsize $x_0^2 x_1$};
\node (120) at (4,0) [inner sep = 1pt] {\scriptsize $x_0 x_1^2$};
\node (030) at (6,0) [inner sep = 1pt] {\scriptsize $x_1^3$};

\draw [<-,very thin] (300) -- (210);
\draw [<-,very thin] (210) -- (120);
\draw [<-,very thin] (120) -- (030);

\node (201) at (2,-1.2) [inner sep = 1pt] {\scriptsize $x_0^2 x_2$};
\node (111) at (4,-1.2) [inner sep = 1pt] {\scriptsize $x_0 x_1 x_2$};
\node (021) at (6,-1.2) [inner sep = 1pt] {\scriptsize $x_1^2 x_2$};

\draw [->,very thin] (201) -- (210);
\draw [->,very thin] (111) -- (120);
\draw [->,very thin] (021) -- (030);

\draw [<-,very thin] (201) -- (111);
\draw [<-,very thin] (111) -- (021);

\node (102) at (4,-2.4) [inner sep = 1pt] {\scriptsize $x_0 x_2^2$};
\node (012) at (6,-2.4) [inner sep = 1pt] {\scriptsize $x_1 x_2^2$};

\draw [->,very thin] (102) -- (111);
\draw [->,very thin] (012) -- (021);

\draw [->,very thin] (012) -- (102);

\node (003) at (6,-3.6) [inner sep = 1pt] {\scriptsize $x_2^3$};

\draw [->,very thin] (003) -- (012);

\node at (0,-4.2) [] {};
\draw [rounded corners,thick] (-1,0.6) -- (5,0.6) -- (5,-1.8) -- (1,-1.8) -- (1,-0.6) -- (-1,-0.6) -- cycle;
\end{tikzpicture}
\qquad\quad
\begin{tikzpicture}[>=latex,scale=0.65]
\node (400) at (-2,1.2) [inner sep = 1pt] {\scriptsize $x_0^4$};
\node (310) at (0,1.2) [inner sep = 1pt] {\scriptsize $x_0^3 x_1$};
\node (220) at (2,1.2) [inner sep = 1pt] {\scriptsize $x_0^2 x_1^2$};
\node (130) at (4,1.2) [inner sep = 1pt] {\scriptsize $x_0 x_1^3$};
\node (040) at (6,1.2) [inner sep = 1pt] {\scriptsize $x_1^4$};

\draw [<-,very thin] (400) -- (310);
\draw [<-,very thin] (310) -- (220);
\draw [<-,very thin] (220) -- (130);
\draw [<-,very thin] (130) -- (040);

\node (300) at (0,0) [inner sep = 1pt] {\scriptsize $x_0^3 x_2$};
\node (210) at (2,0) [inner sep = 1pt] {\scriptsize $x_0^2 x_1 x_2$};
\node (120) at (4,0) [inner sep = 1pt] {\scriptsize $x_0 x_1^2 x_2$};
\node (030) at (6,0) [inner sep = 1pt] {\scriptsize $x_1^3 x_2$};

\draw [->,very thin] (300) -- (310);
\draw [->,very thin] (210) -- (220);
\draw [->,very thin] (120) -- (130);
\draw [->,very thin] (030) -- (040);

\draw [<-,very thin] (300) -- (210);
\draw [<-,very thin] (210) -- (120);
\draw [<-,very thin] (120) -- (030);

\node (201) at (2,-1.2) [inner sep = 1pt] {\scriptsize $x_0^2 x_2^2$};
\node (111) at (4,-1.2) [inner sep = 1pt] {\scriptsize $x_0 x_1 x_2^2$};
\node (021) at (6,-1.2) [inner sep = 1pt] {\scriptsize $x_1^2 x_2^2$};

\draw [->,very thin] (201) -- (210);
\draw [->,very thin] (111) -- (120);
\draw [->,very thin] (021) -- (030);

\draw [<-,very thin] (201) -- (111);
\draw [<-,very thin] (111) -- (021);

\node (102) at (4,-2.4) [inner sep = 1pt] {\scriptsize $x_0 x_2^3$};
\node (012) at (6,-2.4) [inner sep = 1pt] {\scriptsize $x_1 x_2^3$};

\draw [->,very thin] (102) -- (111);
\draw [->,very thin] (012) -- (021);

\draw [->,very thin] (012) -- (102);

\node (003) at (6,-3.6) [inner sep = 1pt] {\scriptsize $x_2^4$};

\draw [->,very thin] (003) -- (012);

\draw [rounded corners, thick] (-3,1.8) -- (7,1.8) -- (7,0.6) -- (5,0.6) -- (5,-1.8) -- (1,-1.8) -- (1,-0.6) -- (-1,-0.6) -- (-1,0.6) -- (-3,0.6) -- cycle;
\end{tikzpicture}
\end{center}
\caption{\label{fig:BorelSet} The Borel sets defined in $\TT_{2,3}$ and $\TT_{2,4}$ by the ideal $(x_0^2,x_0x_1,x_1^4) \subset \KK[x_0,x_1,x_2]$.}
\end{figure}

We need to relate the properties of a strongly stable ideal with its Hilbert polynomial. If $I$ is a strongly stable ideal, for each $s \in \mathbb{N}$ the monomial basis of the homogeneous piece $I_s$ of the ideal is a subset of $\TT_{n,s}$ closed by increasing elementary moves. We call \emph{Borel sets} such subsets of $\TT_{n,s}$ (see Figure \ref{fig:BorelSet} for an example). Proposition \ref{prop:BorelProperties}\emph{(\ref{it:BorelProperties_i})} implies that the monomial basis of $I_s$ for a saturated strongly stable ideal $I \subset \KK[\xx]$ with Hilbert polynomial $p(t)$ and regularity at most $s$ is a Borel set with $q(s) := \binom{n+s}{n} - p(s)$ elements. Thus, we consider the following map
\begin{equation}\label{eq:injective}
\left\{\begin{array}{c} \text{saturated strongly stable ideals in }\KK[\xx] \text{ with}\\   \text{Hilbert polynomial } p(t) \text{ and regularity } \leq s \end{array}\right\} \hookrightarrow \left\{ \begin{array}{c} \text{Borel sets of }\TT_{n,s}\\ \text{with } q(s) \text{ elements}\end{array}\right\}.
\end{equation}
Moreover, Gotzmann's Regularity Theorem suggests to consider $s$ equal to the Gotzmann number of $p(t)$ to determine all saturated strongly stable ideals with Hilbert polynomial $p(t)$. Obviously, there are many Borel sets in $\TT_{n,s}$ with $q(s)$ elements not corresponding to an ideal with Hilbert polynomial $p(t)$. To identify the image of the previous map, we recall a definition and a proposition by Mall.

\begin{definition}[{\cite[Definition 2.7]{Mall1}}] Let $B \subset \TT_{n,s}$ be a Borel set. The set $B^{(i)} := \{ \xx^{\al} \in B\ \vert\ \min \xx^{\al} = n-i \}$ is called \emph{$i$-growth class} of $B$. The sequence $\text{gv}(B) := (\vert B^{(0)}\vert,\ldots,\vert B^{(n)}\vert)$ is called \emph{growth vector} of $B$.
\end{definition}

\begin{proposition}[{\cite[Proposition 3.2]{Mall1}}]
Let $I\subset \KK[\xx]$ be a strongly stable ideal generated by the monomials of a Borel set $B \subset \TT_{n,s}$ and let $p(t)$ be its Hilbert polynomial. Then,
\begin{equation}\label{eq:hilbPolyGV}
p(t) = \binom{n+t}{n} - \sum_{k=0}^n \big\vert B^{(k)}\big\vert \binom{k+t-s}{k},\qquad \forall\ t \geq s.
\end{equation}
\end{proposition}

We can use this result to determine the growth vector of a Borel set $B \subset \TT_{n,s}$ starting from the Hilbert polynomial. The $i$-th difference polynomial of $p(t)$ is
\[
\big(\Delta^i p\big) (t) = \big(\Delta^{i-1} p\big) (t) - \big(\Delta^{i-1} p\big) (t-1) = \binom{n+t-i}{n-i} - \sum_{k=i}^n \big\vert B^{(k)}\big\vert \binom{k+t-s-i}{k-i}.
\]
Evaluating these identities at $t = s$, we obtain the linear system 
\begin{equation}\label{eq:systemGV}
\begin{cases}
\sum_{k=0}^n \big\vert B^{(k)}\big\vert = \binom{n+s}{n} - p(s) \\
\quad \vdots \\
 \sum_{k=i}^n \big\vert B^{(k)}\big\vert = \binom{n+s-i}{n-i} - \big(\Delta^{i} p\big)(s)\\
 \quad\vdots \\
\big\vert B^{(n)}\big\vert = \binom{s}{0} - \big(\Delta^{n} p\big)(s) 
 \end{cases}
\end{equation}
whose solution is
\[
 \big\vert B^{(i)} \big\vert =  \sum_{k=i}^n \big\vert B^{(k)}\big\vert - \sum_{k=i+1}^n \big\vert B^{(k)}\big\vert = \binom{n+s-i-1}{n-i} - \big(\Delta^i p\big)(s) + \big(\Delta^{i+1}p\big)(s), \qquad i < n
\]
 and $\vert B^{(n)}\vert = 1$ (recall that $\big(\Delta^{i} p\big)(t) \equiv 0$ for $i > \deg p(t)$ and $\deg p(t) < n$).
Let us call  \emph{growth vector} of $p(t)$ in degree $s$ the solution of the linear system \eqref{eq:systemGV} and let us denote it by 
$\text{gv}_s\big(p(t)\big)$.
 
\begin{proposition}[{cf.~\cite[Theorem 3.3]{LellaBorel}}]\label{prop:main}
Let $p(t)$ be a Hilbert polynomial. There is a bijective map
\begin{equation}\label{eq:identification}
\begin{array}{ccccc}
\medskip
\left\{\begin{array}{c} \text{saturated strongly stable ideals} \\ \text{in }\KK[\xx] \text{ with Hilbert polynomial} \\ p(t) \text{ and regularity } \leq s \end{array}\right\} && \xleftrightarrow{1:1} && \left\{ \begin{array}{c} \text{Borel sets of }\TT_{n,s}\\ \text{with } q(s) \text{ elements and}\\ \text{growth vector } \textnormal{gv}_{s}\big(p(t)\big)\end{array}\right\}\\
\smallskip
 I && \longrightarrow && \text{monomial basis of } I_s \\
 \text{saturation of } (B) && \longleftarrow && B
\end{array}
\end{equation}
\end{proposition}

In order to determine the Borel sets of Proposition \ref{prop:main}, we use a recursive algorithm based on Proposition \ref{prop:BorelProperties}\emph{(\ref{it:BorelProperties_iii})}. Indeed, if $I \subset \KK[x_0,\ldots,x_n]$ is a strongly stable ideal with Hilbert polynomial $p(t)$ and $B$ is the associated Borel set in $\TT_{n,s}$, then the subset $B' = \{\xx^{\al} \in B\ \vert\ \min \xx^{\al} > n\} \subset B$ is a Borel set in $\TT_{n-1,s}$ corresponding to the strongly stable ideal $I' = (x_n,I) \cap \KK[x_0,\ldots,x_{n-1}] \subset \KK[x_0,\ldots,x_{n-1}]$ with Hilbert polynomial $\big(\Delta p\big)(t)$.

\begin{figure}
\begin{center}
\begin{tikzpicture}[>=latex,scale=0.6]

\node (500) at (-4,2.4) [inner sep = 1pt] {\tiny $x_0^5$};
\node (410) at (-2,2.4) [inner sep = 1pt] {\tiny $x_0^4 x_1$};
\node (320) at (0,2.4) [inner sep = 1pt] {\tiny $x_0^3 x_1^2$};
\node (230) at (2,2.4) [inner sep = 1pt] {\tiny $x_0^2 x_1^3$};
\node (140) at (4,2.4) [inner sep = 1pt] {\tiny $x_0 x_1^4$};
\node (050) at (6,2.4) [inner sep = 1pt] {\tiny $x_1^5$};

\draw [<-,very thin] (500) -- (410);
\draw [<-,very thin] (410) -- (320);
\draw [<-,very thin] (320) -- (230);
\draw [<-,very thin] (230) -- (140);
\draw [<-,very thin] (140) -- (050);

\node (400) at (-2,1.2) [inner sep = 1pt] {\tiny $x_0^4 x_2$};
\node (310) at (0,1.2) [inner sep = 1pt] {\tiny $x_0^3 x_1 x_2$};
\node (220) at (2,1.2) [inner sep = 1pt] {\tiny $x_0^2 x_1^2 x_2$};
\node (130) at (4,1.2) [inner sep = 1pt] {\tiny $x_0 x_1^3 x_2$};
\node (040) at (6,1.2) [inner sep = 1pt] {\tiny $x_1^4 x_2$};

\draw [<-,very thin] (410) -- (400);
\draw [<-,very thin] (320) -- (310);
\draw [<-,very thin] (230) -- (220);
\draw [<-,very thin] (140) -- (130);
\draw [<-,very thin] (050) -- (040);

\draw [<-,very thin] (400) -- (310);
\draw [<-,very thin] (310) -- (220);
\draw [<-,very thin] (220) -- (130);
\draw [<-,very thin] (130) -- (040);

\node (300) at (0,0) [inner sep = 1pt] {\tiny $x_0^3 x_2^2$};
\node (210) at (2,0) [inner sep = 1pt] {\tiny $x_0^2 x_1 x_2^2$};
\node (120) at (4,0) [inner sep = 1pt] {\tiny $x_0 x_1^2 x_2^2$};
\node (030) at (6,0) [inner sep = 1pt] {\tiny $x_1^3 x_2^2$};

\draw [->,very thin] (300) -- (310);
\draw [->,very thin] (210) -- (220);
\draw [->,very thin] (120) -- (130);
\draw [->,very thin] (030) -- (040);

\draw [<-,very thin] (300) -- (210);
\draw [<-,very thin] (210) -- (120);
\draw [<-,very thin] (120) -- (030);

\node (201) at (2,-1.2) [inner sep = 1pt] {\tiny $x_0^2 x_2^3$};
\node (111) at (4,-1.2) [inner sep = 1pt] {\tiny $x_0 x_1 x_2^3$};
\node (021) at (6,-1.2) [inner sep = 1pt] {\tiny $x_1^2 x_2^3$};

\draw [->,very thin] (201) -- (210);
\draw [->,very thin] (111) -- (120);
\draw [->,very thin] (021) -- (030);

\draw [<-,very thin] (201) -- (111);
\draw [<-,very thin] (111) -- (021);

\node (102) at (4,-2.4) [inner sep = 1pt] {\tiny $x_0 x_2^4$};
\node (012) at (6,-2.4) [inner sep = 1pt] {\tiny $x_1 x_2^4$};

\draw [->,very thin] (102) -- (111);
\draw [->,very thin] (012) -- (021);

\draw [->,very thin] (012) -- (102);

\node (003) at (6,-3.6) [inner sep = 1pt] {\tiny $x_2^5$};

\draw [->,very thin] (003) -- (012);

\draw [rounded corners, thick] (-5,3) -- (5,3) -- (5,1.8) -- (3,1.8) -- (3,-0.6) -- (-1,-0.6) -- (-1,0.6) -- (-3,0.6) -- (-3,1.8) -- (-5,1.8) -- cycle;

\end{tikzpicture}
\quad
%
%
%
%
%
\quad
\begin{tikzpicture}[>=latex,scale=0.6]

\node (500) at (-4,2.4) [inner sep = 1pt] {\tiny $x_0^5$};
\node (410) at (-2,2.4) [inner sep = 1pt] {\tiny $x_0^4 x_1$};
\node (320) at (0,2.4) [inner sep = 1pt] {\tiny $x_0^3 x_1^2$};
\node (230) at (2,2.4) [inner sep = 1pt] {\tiny $x_0^2 x_1^3$};
\node (140) at (4,2.4) [inner sep = 1pt] {\tiny $x_0 x_1^4$};
\node (050) at (6,2.4) [inner sep = 1pt] {\tiny $x_1^5$};

\draw [<-,very thin] (500) -- (410);
\draw [<-,very thin] (410) -- (320);
\draw [<-,very thin] (320) -- (230);
\draw [<-,very thin] (230) -- (140);
\draw [<-,very thin] (140) -- (050);

\node (400) at (-2,1.2) [inner sep = 1pt] {\tiny $x_0^4 x_2$};
\node (310) at (0,1.2) [inner sep = 1pt] {\tiny $x_0^3 x_1 x_2$};
\node (220) at (2,1.2) [inner sep = 1pt] {\tiny $x_0^2 x_1^2 x_2$};
\node (130) at (4,1.2) [inner sep = 1pt] {\tiny $x_0 x_1^3 x_2$};
\node (040) at (6,1.2) [inner sep = 1pt] {\tiny $x_1^4 x_2$};

\draw [<-,very thin] (410) -- (400);
\draw [<-,very thin] (320) -- (310);
\draw [<-,very thin] (230) -- (220);
\draw [<-,very thin] (140) -- (130);
\draw [<-,very thin] (050) -- (040);

\draw [<-,very thin] (400) -- (310);
\draw [<-,very thin] (310) -- (220);
\draw [<-,very thin] (220) -- (130);
\draw [<-,very thin] (130) -- (040);

\node (300) at (0,0) [inner sep = 1pt] {\tiny $x_0^3 x_2^2$};
\node (210) at (2,0) [inner sep = 1pt] {\tiny $x_0^2 x_1 x_2^2$};
\node (120) at (4,0) [inner sep = 1pt] {\tiny $x_0 x_1^2 x_2^2$};
\node (030) at (6,0) [inner sep = 1pt] {\tiny $x_1^3 x_2^2$};

\draw [->,very thin] (300) -- (310);
\draw [->,very thin] (210) -- (220);
\draw [->,very thin] (120) -- (130);
\draw [->,very thin] (030) -- (040);

\draw [<-,very thin] (300) -- (210);
\draw [<-,very thin] (210) -- (120);
\draw [<-,very thin] (120) -- (030);

\node (201) at (2,-1.2) [inner sep = 1pt] {\tiny $x_0^2 x_2^3$};
\node (111) at (4,-1.2) [inner sep = 1pt] {\tiny $x_0 x_1 x_2^3$};
\node (021) at (6,-1.2) [inner sep = 1pt] {\tiny $x_1^2 x_2^3$};

\draw [->,very thin] (201) -- (210);
\draw [->,very thin] (111) -- (120);
\draw [->,very thin] (021) -- (030);

\draw [<-,very thin] (201) -- (111);
\draw [<-,very thin] (111) -- (021);

\node (102) at (4,-2.4) [inner sep = 1pt] {\tiny $x_0 x_2^4$};
\node (012) at (6,-2.4) [inner sep = 1pt] {\tiny $x_1 x_2^4$};

\draw [->,very thin] (102) -- (111);
\draw [->,very thin] (012) -- (021);

\draw [->,very thin] (012) -- (102);

\node (003) at (6,-3.6) [inner sep = 1pt] {\tiny $x_2^5$};

\draw [->,very thin] (003) -- (012);

\draw [rounded corners, thick] (-5,3) -- (5,3) -- (5,0.6) -- (1,0.6) -- (1,-0.6) -- (-1,-0.6) -- (-1,0.6) -- (-3,0.6) -- (-3,1.8) -- (-5,1.8) -- cycle;

\end{tikzpicture}
%
%
\end{center}
\caption{\label{fig:example2} Borel sets in $\TT_{2,5}$ corresponding to the saturated strongly stable ideals $(x_0^3,x_0^2 x_1,x_0 x_1^4)$ (on the left) and $(x_0^3,x_0^2 x_1^2,x_0 x_1^3)$ (on the right) in 3 variables with Hilbert polynomial $t+6$ and regularity at most 5.}
\end{figure}
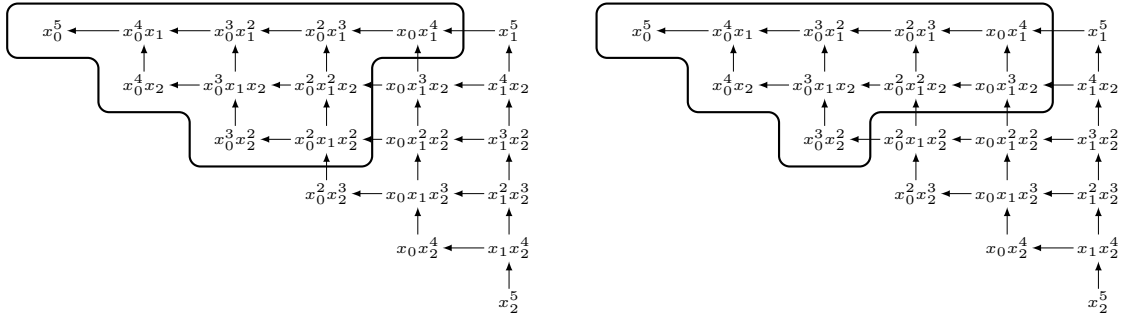

\begin{example} We want to determine the set of strongly stable ideals in the polynomial ring $\KK[x_0,x_1,x_2]$ with regularity at most 5 defining schemes with Hilbert polynomial $p(t) = t+6$. The Gotzmann number of $p(t)$ is $6$ and its growth vector in degree $5$ is $\text{gv}_5(t+6) = (5,4,1)$. We start considering the set of strongly stable ideals in $\KK[x_0,x_1]$ with Hilbert polynomial $\Delta p(t) = 1$ and regularity at most $5$ corresponding to Borel sets with growth vector $\text{gv}_{5}(\Delta p(t)) = (4,1)$. There is a unique Borel set $B' = \left\{x_0^5,x_0^4 x_1, x_0^3 x_1^2, x_0^2 x_1^3, x_0 x_1^4\right\}$. Since $x_1^5$ is not contained in $B'$, a Borel set $B \subset \TT_{2,5}$ with growth vector $(5,4,1)$ does not contain monomials obtained from $x_1^5$ by applying decreasing elementary moves, i.e.~$x_1^4 x_2$, $x_1^3 x_2^2$, $x_1^2 x_2^3$, $x_1 x_2^4$ and $x_2^5$. Hence, we need to select 5 monomials divisible by both $x_0$ and $x_2$ producing a set closed by increasing elementary moves (see Figure \ref{fig:example2}).
\end{example}

This package provides the method \texttt{stronglyStableIdeals} to compute the set of strongly stable ideals of a given polynomial ring with fixed Hilbert polynomial and bounded regularity.

\medskip

\begin{Verbatim}
i9 : stronglyStableIdeals (4*t, QQ[x,y,z,w])

                 5   4 2                     2   4    5                2     2   4                2   3
o9 = {ideal (x, y , y z ), ideal (x*z, x*y, x , y z, y ), ideal (x*y, x , x*z , y ), ideal (x*y, x , y )}

o9 : List

i10 : stronglyStableIdeals (4*t, QQ[x,y,z,w], MaxRegularity => 4)

                    2     2   4                2   3
o10 = {ideal (x*y, x , x*z , y ), ideal (x*y, x , y )}

o10 : List
\end{Verbatim}

\section{Segment ideals}

The transitive closure of the order relation
\begin{equation}\label{eq:BorelOrder}
 \xx^{\al} >_B \xx^{\boldsymbol{\beta}} \quad \Longleftrightarrow\quad \xx^{\boldsymbol{\beta}} = \down{i}(\xx^{\al})
\end{equation}
induces a partial order on the set of monomials of any degree called \emph{Borel order}. Every graded term ordering is a refinement of this partial order. Since a Borel set $B$ is closed with respect to the Borel order, i.e.~$\xx^{\al} >_B \xx^{\boldsymbol{\beta}},\ \xx^{\boldsymbol{\beta}} \in B \Rightarrow \xx^{\al} \in B$, it is natural to ask whether there exists a term ordering $\prec$ with the same property. For instance, for the lexicographic ideal, the graded lexicographic order separates, in each degree, monomials contained in the ideal from those outside. In \cite{CLMR} several notions of segment ideals are introduced.

\begin{definition}[{\cite[Definition 3.1, Definition 3.7]{CLMR}}]
A Borel set $B \subset \TT_{n,s}$ is called \emph{segment} if there exists a term ordering $\prec$ such that $\xx^{\al} \succ \xx^{\boldsymbol{\beta}}$, for all $\xx^{\al} \in B$ and $\xx^{\boldsymbol{\beta}} \in \TT_{n,s}\setminus B$. 

Let $I \subset \KK[\xx]$ be a saturated strongly stable ideal.
\begin{enumerate}[(i)]
\item $I$ is called \emph{hilb-segment} if the Borel set $I \cap \TT_{n,r}$ is a segment, where $r$ is the Gotzmann number of the Hilbert polynomial of $I$.
\item $I$ is called \emph{reg-segment} if the Borel set $I \cap \TT_{n,m}$ is a segment, where $m$ is the regularity of $I$.
\item $I$ is called \emph{gen-segment} if there exists a term ordering $\prec$ such that $\xx^\alpha \succ \xx^{\boldsymbol{\beta}}$ for each minimal generator $\xx^{\al}$ of degree $s$ of $I$ and for all $\xx^{\boldsymbol{\beta}} \in \TT_{n,s} \setminus I_s$.
\end{enumerate}
\end{definition}

These notions are very important in the construction of flat families based on properties of Gr\"obner bases and in general for the study of the Hilbert scheme. The package provides three methods for determining whether a strongly stable ideal may be some type of segment (and, in case, gives the term ordering). These methods use tools of the package \texttt{gfanInterface} and the term ordering is given as a weight vector.

\medskip

\begin{Verbatim}
i11 : sevenPointsP2 = stronglyStableIdeals (7, 3, MaxRegularity => 5)

               2     2   5           2   4     3             2   2     3   4
o11 = {ideal (x , x x , x ), ideal (x , x , x x ), ideal (x x , x x , x , x )}
               0   0 1   1           0   1   0 1           0 1   0 1   0   1

o11 : List

i12 : for J in sevenPointsP2 list isHilbSegment J

o12 = {(true, {7, 3, 1}), (false, ), (true, {4, 3, 1})}

o12 : List

i13 : for J in sevenPointsP2 list isRegSegment J

o13 = {(true, {7, 3, 1}), (false, ), (true, {4, 3, 1})}

o13 : List

i14 : for J in sevenPointsP2 list isGenSegment J

o14 = {(true, {6, 3, 1}), (true, {4, 3, 1}), (true, {4, 3, 1})}

o14 : List
\end{Verbatim}

\providecommand{\bysame}{\leavevmode\hbox to3em{\hrulefill}\thinspace}
\providecommand{\MR}{\relax\ifhmode\unskip\space\fi MR }
\providecommand{\MRhref}[2]{%
  \href{http://www.ams.org/mathscinet-getitem?mr=#1}{#2}
}
\providecommand{\href}[2]{#2}

\bigskip

\end{document}